\documentclass{article}

 \PassOptionsToPackage{numbers, compress}{natbib}
\usepackage[preprint]{neurips_2020}

\usepackage[utf8]{inputenc} % allow utf-8 input
\usepackage[T1]{fontenc}    % use 8-bit T1 fonts
\usepackage{hyperref}       % hyperlinks
\usepackage{graphicx}
\usepackage{url}            % simple URL typesetting
\usepackage{booktabs}       % professional-quality tables
\usepackage{amsfonts}       % blackboard math symbols
\usepackage{nicefrac}       % compact symbols for 1/2, etc.
\usepackage{microtype}      % microtypography
\usepackage{algorithm}  
\usepackage{algpseudocode}  
\usepackage{amsmath}  
\title{Federated Two-stage Learning with Sign-based Voting}

\author{%
  Zichen Ma\\
  CUHK(SZ)\\
  \And
  Zihan Lu \\
  Pingan Technology\\
  \And
  Yu Lu \\
  CUHK(SZ)\\
  \And
  Wenye Li \\
  CUHK(SZ)\\
  \And
  Jinfeng Yi \\
  JD AI\\
  \And
  Shuguang Cui \\
  CUHK(SZ)\\
}

\begin{document}

\maketitle

\begin{abstract}
Federated learning is a distributed machine learning mechanism where local devices collaboratively train a shared global model under the orchestration of a central server, while keeping all private data decentralized. In the system, model parameters and its updates are transmitted instead of raw data, and thus the communication bottleneck has become a key challenge. Besides, recent larger and deeper machine learning models also pose more difficulties in deploying them in a federated environment. In this paper, we design a federated two-stage learning framework that augments prototypical federated learning with a cut layer on devices and uses sign-based stochastic gradient descent with the majority vote method on model updates. Cut layer on devices learns informative and low-dimension representations of raw data locally, which helps reduce global model parameters and prevents data leakage. Sign-based SGD with the majority vote method for model updates also helps alleviate communication limitations. Empirically, we show that our system is an efficient and privacy preserving federated learning scheme and suits for general application scenarios.
\end{abstract}

\section{Introduction}
Federated learning is a distributed machine learning mechanism where local institutions or devices collaboratively train a shared global model under the orchestration of a central server, while keeping all the sensitive private data decentralized \cite{kairouz2019advances}. Challenges in federated learning include an unbalanced and non-IID (identically and independently distributed) data allocation on an enormous number of devices \cite{moreno2012unifying} and limited communication bandwidth \cite{zhang2013information}.

The recent deeper and larger machine learning models \cite{devlin2018bert} violate the limitation of communication channels because traditional federated learning trains a shared global model via communicating parameters and its updates to each device \cite{konevcny2016federated}. This needs a new paradigm other than the prototypical federated learning framework. In this paper, we design a federated two-stage learning framework that augments prototypical federated learning with a cut layer on devices and uses sign-based stochastic gradient descent with the majority vote method on model updates. Devices with a cut layer split the execution of a model on a per-layer basis, which can help learn informative and compact representations of raw data (smashed data) locally. The global model is then trained using the SIGNSGD algorithm \cite{riedmiller1993direct} based on the low-dimension smashed data from devices. 

%\paragraph{Our contribution} (\romannumeral1) Efficiency: By splitting the execution of a model between the devices and the server, local devices learn informative and low-dimension representations of raw data which means the global model needs less parameters and thus reduces the required communicating ones. In addition, SIGNSGD with majority vote \cite{bernstein2018signsgd} has been proved that can alleviate communication bottleneck by transmitting just the sign of each minibatch stochastic gradient while preserving competitive results on publicly available real-world datasets.

Compared with the existing approaches, our proposed model has several advantages. First, the proposed federated learning scheme is highly  efficient. By splitting the execution of a model between devices and the server, local devices learn informative and low-dimension representations of raw data. The global model then needs fewer parameters because of these compact inputs and thus reduces the required communicated intermediaries. Besides, SIGNSGD with the majority vote \cite{bernstein2018signsgd} has been proved that can alleviate the communication bottleneck by transmitting just the sign of each minibatch stochastic gradients while preserving competitive results on publicly available real-world datasets.

Second, our designed framework suits for general application scenarios. One assumption in traditional federated learning is that all the data is from the same source (i.e., text \cite{hard2018federated} or images), which may not be realistic in real-world scenarios. Note that devices may contain multiple sources of data such as videos, images, or text. A single global model may not be able to handle all of them. Besides, it is not likely to infer using the trained model when new devices contain different data modalities that have not been observed during the training process.  The proposed model, on the contrary, can handle data depending on their sources and distributions. Our model is capable of handling different sources of data that have not been observed before and thus suits for general applications.

In addition, the proposed scheme is a privacy preserving federated learning architecture. Intermediaries communicated between devices and the server in the traditional federated learning may still leak some information about raw data. This leakage does harm to the constraint that raw data must remain private. Our proposed federated two-stage learning mechanism, however, reduces invertibility of intermediate representations by minimizing distance correlation \cite{szekely2007measuring} between the smashed data and the raw data while still ensuring model's prediction accuracy. Experiments show that our model can yield superior or comparable results with the state-of-the-art methods with less leakage of sensitive information.

%% to be modified
% The paper is organized as follows. First we introduce the related work in Section 2, and present our model in Section 3. After reporting the empirical results in Section 4, we conclude the paper with discussion and future work in Section 5.

\section{Related Work}

\begin{figure}[t]
\centering
\includegraphics[scale=0.5]{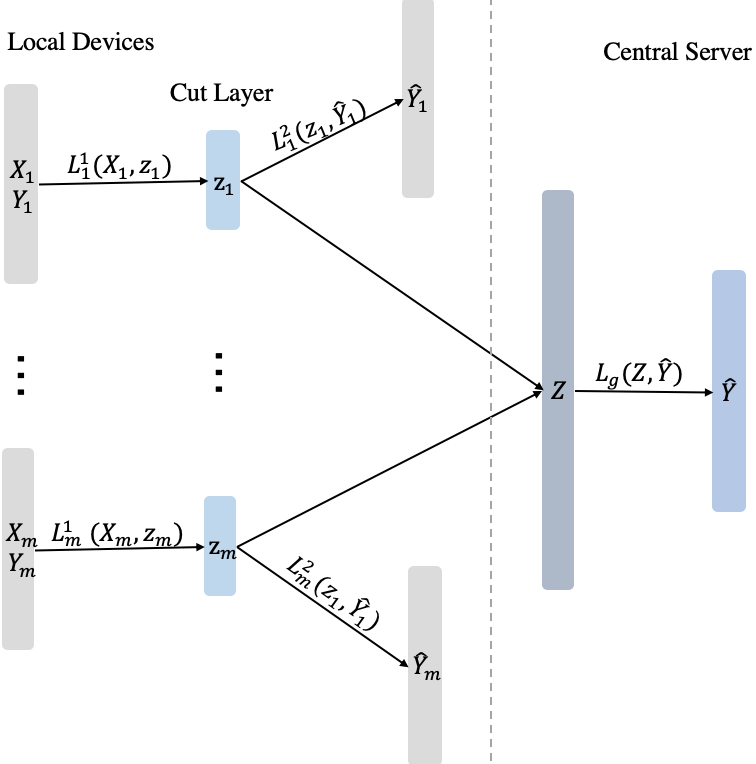}
\caption{An overview of the proposed federated two-stage learning with sign-based voting system}
\label{Figure1}
\end{figure}

\paragraph{Federated Learning} Federated learning enables multiple parties to collaboratively build a machine learning model while keeping their data private \cite{yang2019federated}. Challenges in federated learning include unbalanced and non-IID data partition, unreliable devices and limited communication bandwidth. The most popular optimization method in federated learning is the Federated Averaging algorithm \cite{mcmahan2016communication} which aggregates local models in a manner of weighted average. Li et al. \cite{li2018federated} added a proximal term to the objective to improve the stability of algorithm. Fully decentralized learning tries to alleviate trust concerns about server \cite{vanhaesebrouck2017decentralized}. Caldas et al. \cite{caldas2018expanding} proposed an algorithm that improves the efficiency of federated learning. 
%In this paper, we design Federated Split learning with sign-based voting to improve efficiency, universality and security of federated learning.

%paragraph{Distributed Deep Learning Methods} Split learning \cite{gupta2018distributed} is a resource efficient distributed learning method using intermediate representations of the cut layer and transmitting them to another entity to train models. The key idea behind split learning is to split the execution of a model between clients and servers \cite{kairouz2019advances}. %Our designed method reduces the potential data leakage in communications by minimizing distance correlation between smashed and raw data and improves security of the system.

\paragraph{Compressed Optimization} Training large models in distributed learning setting requires high communication cost \cite{liu2010distributed} and thus many compressed optimization methods are proposed to alleviate this challenge. SIGNSGD tries to transmit the sign of each minibatch stochastic gradients instead of gradients themselves. It has been proved that SIGNSGD performs better than traditional SGD in efficiency \cite{bernstein2018signsgd}.

\section{Federated Two-stage Learning}

Prototypical federated learning builds a global model on the central server and transmits model parameters to local devices. Then each local device computes gradient updates based on its own subset of the high-dimension raw data. After that, model parameters are updated using gradients. Then these updated parameters are transmitted back to the server where all updates are aggregated in a manner of weighted average. Finally, the global model is updated and a new round is started until the model converges \cite{yang2019federated}. This process uses high-dimension raw data as the global model's local input, which requires many intermediaries (e.g. model parameters and its updates) to be transmitted in multiple rounds, which poses high communication cost in the system.

Our proposed federated two-stage learning, on the contrary, divides the execution of a model on a per-layer basis between devices and the server. The designed method can be adopted for both training and inference. An overview of our system is shown in Figure\ref{Figure1}. Devices try to learn informative and compact representations of the raw data locally. Then the shared global model will directly use these representations as input rather than the raw data. 

The designed system is divided into two complementary parts: local learning and global aggregation. The first process aims to extract representative and low-dimension features from the raw data that are important for training the model. Then the global model can operate on these compact features and thus reduce its parameters. 

To illustrate how the system works, first we introduce local operations in the federated two-stage learning with sign-based voting. Then we demonstrate how to reduce global model parameters and eliminate potential data leakage. After showing the global model aggregation process and providing the pseudo-code of the optimization algorithm, we describe how to infer using the trained model in system.

\subsection{Local Split Learning} 

For each device with a subset of data $(X_k, Y_k)$, it learns a representative and low-dimension feature $z_k$, which is also known as smashed data. There are many discussions on the choice of smashed data \cite{sharma2019expertmatcher}, but in general the following properties should be satisfied: (\romannumeral1) it must be representative that captures the important features in the raw data $X_k$, (\romannumeral2) it should be compact enough comparing with $X_k$, (\romannumeral3) it cannot be inverted back to $X_k$ easily because preserving data privacy is a key concern in federated learning. 

In the paper, we find smashed data by minimizing the logarithm of distance correlation (DCOR) between the smashed data and the raw data. We show that minimizing DCOR minimizes their Kullback-Leibler divergence, which is a measure of invertibility of the smashed data in information theory. For simplicity, we use distance covariance (DCOV) which is an unnormalized DCOR \cite{vepakomma2019reducing}, Kullback-Leibler divergence $D_{KL}$ and cross entropy $H$ to build the connection. 

From Vepakomma et al. \cite{vepakomma2018supervised}, the sample DCOV can be derived from covariance matrices $Cov(X_k)$, $Cov(z_k)$.
\begin{equation}
\label{E1}
    \begin{aligned}
        DCOV(X_k, z_k) &= Tr(X_k^TX_kz_k^Tz_k)\\
          &= n^2Tr(Cov(X_k)Cov(z_k))
    \end{aligned}
\end{equation}
According to arithmetic geometric mean inequality \cite{bhatia1993more}, we have,
\begin{equation}
\label{E2}
Tr(Cov(X_k)Cov(z_k)) \geq det(Cov_{z_kX_k})det(Cov_{X_kz_k})
\end{equation}
In equation (\ref{E2}), $Cov_{z_kX_k}$ is the cross-covariance matrix and $det(Cov_{z_kX_k})$ is the cross entropy $H(X_k, z_k)$. The KL divergence relates to cross entropy as 
\begin{equation}
\label{E3}
D_{KL}(X_k||z_k) = H(X_k, z_k) - H(X_k)
\end{equation}
Therefore we combine equations (\ref{E1})(\ref{E2})(\ref{E3}) and build the connection between DCOR and KL divergence. Moreover, we prove that minimizing DCOR also minimizes the invertibility of the smashed data. So we have the first part of local loss,
\begin{equation}
\label{E4}
L_{k}^1 = \log(DCOR(X_k, z_k))
\end{equation}
The first loss $L_{k}^1$, however, just satisfies one of properties proposed above. The smashed data should also be informative which means it extracts important features in $(X_k, Y_k)$ and thus we have local supplementary loss which measures prediction ability of the smashed data,
\begin{equation}
\label{E5}
L_{k}^2 = CE(z_k, \hat{Y}_k)
\end{equation}
The equations (\ref{E4}) and (\ref{E5}) above allow us to find smashed data from local devices efficiently. Given suitable hyper-parameters, we can train the local model using sign-based stochastic gradient descent and the local training objective optimizes model parameters with respect to the local loss function,
\begin{equation}
\label{E6}
L_{local} = \sum_{k=1}^m(\alpha_1L_{k}^1 + \alpha_2L_{k}^2)
\end{equation}
where the system contains $m$ devices. $\alpha_1$ and $\alpha_2$ are hyper-parameters that balance the leakage and precision of the smashed data.

\subsection{Global Model Aggregation} 

Traditional federated learning server directly exchanges global model parameters and its updates with local devices, which depletes communication bandwidth badly because of high-dimension raw data. Key differences between prototypical federated learning and our proposed mechanism include (\romannumeral1) Our global model now operates on the learnt representative and low-dimension smashed data. Therefore, the global model in federated two-stage learning can be much smaller than the traditional federated learning when training on the same objective. Moreover, the number of communicated intermediaries reduces that helps our system be more efficient than prototypical federated learning. (\romannumeral2) Our sign-based optimization scheme just exchanges the sign of each minibatch stochastic gradients in the system. This method compresses client-server communication and still maintains competitive results as other methods.

\begin{algorithm}[tb]
  \caption{Federated Two-stage Learning with Sign-based Voting. The m devices are indexed by k; B is the local minibatch size, C is the participation level and $\delta$ is the learning rate.}  
  \label{algorithm1}  
  \textbf{Server executes:}
  \begin{algorithmic}  
    \State initialize $\omega_0$, sign$\left[\sum_{k} sign(\hat{g}_k)\right]$
    \For{each round $t$ = 1,2, ...}
      \State n $\gets$ $\max(C \cdot m, 1)$
      \State $S_t$ $\gets$ (random set of n clients)
      \For{each client k $\in S_t$}
      \State sign($\hat{g}_k$) $\gets$ ClientUpdate(k)
      \EndFor
      \State return sign$\left[\sum_{k} sign(\hat{g}_k)\right]$ to clients
     \EndFor  
  \end{algorithmic}  
  \textbf{ClientUpdate(k):}
  \begin{algorithmic} 
    \State $B$ $\gets$ split local data into batches of size B
    \For{batch b $\in B$}
    \State $\omega = \omega - \delta$ sign$\left[\sum_{k} sign(\hat{g}_k)\right]$
    \State sign($\hat{g}_k$) = $\nabla L(\omega, b)$
    \EndFor
    \State return sign($\hat{g}_k$) to the server
  \end{algorithmic}  
\end{algorithm}

In the paper, the global model trains iteratively. The server sends the global model parameters and initial sign to each device at the beginning of the first iteration. Each device then runs their own local model and the transmitted global model to predict labels. After that, losses and gradients can be computed between predicted label and true label on their own subset of data. Given a proper loss function, we have the global loss,
\begin{equation}
\label{E7}
L_{global} = \sum_{k=1}^m(L_g(z_k, \hat{Y}))
\end{equation}
And the total loss is a combination of local loss and global loss,
\begin{equation}
\label{E8}
    \begin{aligned}
        L &= L_{local} + \lambda L_{global}\\
          &= \sum_{k=1}^m(\alpha_1L_{k}^1 + \alpha_2L_{k}^2 + \lambda L_g(z_k, \hat{Y}))
    \end{aligned}
\end{equation}
where $\alpha_1$, $\alpha_2$ and $\lambda$ are hyper-parameters that need fine tuning. $\alpha_1$ and $\alpha_2$ control the data leakage and prediction ability of the smashed data and $\lambda$ balances the updates to local model. In the overall loss function, global objective acts as a regularization term to synchronize the local smashed data on devices. In addition, it punishes the local objective in case of overfitting.

\begin{table}[tb]
\caption{Theoretical efficiency comparison between our method and other federated learning algorithms. The proposed Federated Two-stage exchanges sign of gradients in the system rather than model parameters.}
\label{Table2}
\centering
\begin{tabular}{ccc} % 控制表格的格式
\toprule
Method & \# Inter./Device & Total Inter. \\
    \midrule
    Federated Two-stage & $2G$ & $n_1(mN_1+2mG)$\\
    FedAvg \& FedProx & $2 N_2$ & $2 n_2 m N_2$\\
\bottomrule
\end{tabular}
\end{table}

\begin{table}[tb]
\caption{Efficiency comparison between Federated Two-stage, FedAvg and FedProx using MNIST (non-IID), CIFAR-10 (IID) and Shakespeare(non-IID). Our method maintains competitive performance but needs less ($\sim$10$\%$) communication intermediaries. we report average value of 50 runs}
\label{Table1}
\centering
\begin{tabular}{lcccc} % 控制表格的格式
\toprule
Dataset & Method & Test Accuracy & \# Comm. Rounds& Total Comm. \\
    \midrule
      & Federated Two-stage & $97.93$ & 437 &$4.526 \times 10^9$ \\
     MNIST & FedAvg & $98.21$ & 689 & $2.292 \times 10^{10}$ \\
      & FedProx & $98.31$ & 651 & $2.166 \times 10^{10}$\\
    \cline{1-5}  % 这部分是画一条横线在2-6 排之间
     & Federated Two-stage & $81.46$ & 191 &$3.258 \times 10^8$ \\
     CIFAR-10 & FedAvg & $81.95$ & 279 & $2.559 \times 10^{9}$ \\
      & FedProx & $82.14$ & 253 & $2.321 \times 10^{9}$\\
    \cline{1-5}
    & Federated Two-stage & $60.97$ & 744 &$1.694 \times 10^9$ \\
     Shakespeare & FedAvg & $61.52$ & 936 & $1.703 \times 10^{10}$ \\
      & FedProx & $61.28$ & 962 & $1.750 \times 10^{10}$\\
\bottomrule
\end{tabular}
\end{table}

The most common approach to optimizing federated learning is the Federated Averaging algorithm \cite{mcmahan2016communication}. Each device runs some SGD steps locally and then sends the updated local model back to the server. The coordinating server aggregates the local model in a manner of weighted average and finally transmits the updated global model to devices. However, this algorithm communicates model parameters in 32-bit floating point number that consumes lots of resources and is not an efficient method. In our system, we use 1-bit compressed gradient (sign of the gradient) which reduces communication cost a lot. In an iteration, gradients can be computed based on the loss between the predicted label and the true label. Then devices send the sign of its local gradients back to the server. After receiving the sign of gradients from local devices, the server aggregates these gradients by using the majority vote scheme and pushes 1-bit decision back to every devices. Finally, each device updates its model of the form
\begin{equation}
\label{E9}
\omega_{k}^{t+1} = \omega_{k}^{t} - \delta sign\left[\sum_{k=1}^m sign(\hat{g}_k)\right]
\end{equation}
where $\omega$ is the model parameters, $\hat{g}_k$ is the gradients of $k$'s device and $\delta$ is the learning rate. Complete pseudo-code is shown in Algorithm \ref{algorithm1}. 
\subsection{New Devices Inference} 
Prototypical federated learning mechanism often uses Federated Averaging algorithm that is described above. Since it only trains a global model, the new device with $X_{m+1}$ directly makes prediction using the trained model. Comparatively, our proposed method divides the original global model into two parts, so a new device with $X_{m+1}$ needs to be processed first in the local model to learn the informative representations of its raw data. Then the low-dimension smashed data are passed to the trained global model to predict. In the paper, we average all trained local models' logits (the vector of non-normalized predictions) when dealing with new devices. Because new devices may contain unknown data sources and distributions, we cannot simply pass them to any local model. The ensemble approach we proposed shows competitive results in the experiment compared with Federated Averaging algorithm.

\begin{figure*}[tb]
\centering 
\includegraphics[width=4.5cm]{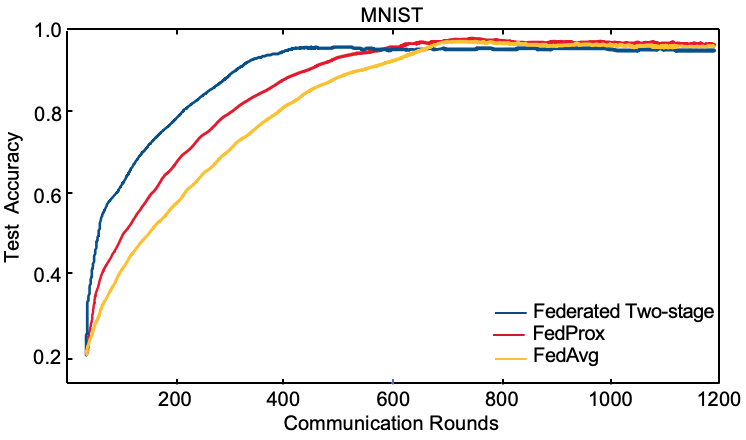}  
\includegraphics[width=4.5cm]{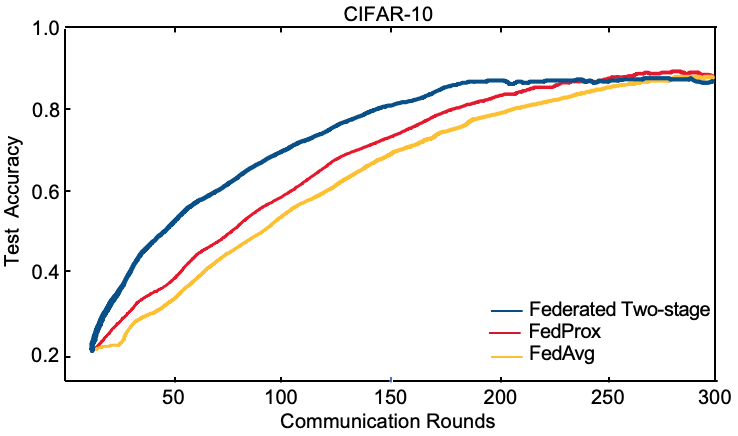}  
\includegraphics[width=4.5cm]{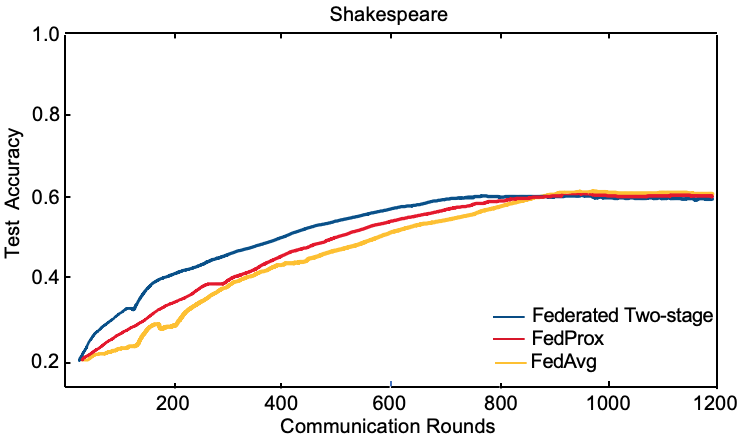}  
\caption{Performance of Federated Two-stage, FedProx and FedAvg algorithm on MNIST, CIFAR-10 and Shakespeare dataset. Federated Two-stage maintains competitive results but uses less communication rounds.}
\label{Figure2}
\end{figure*}

\begin{figure*}[tb]
\centering  
\includegraphics[width=4.5cm]{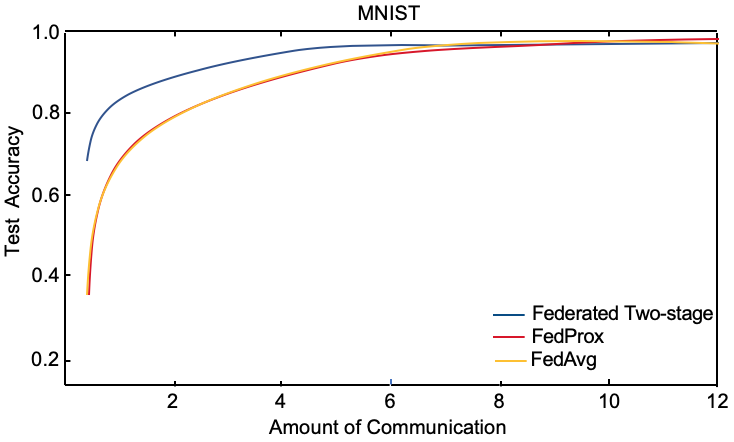}
\includegraphics[width=4.5cm]{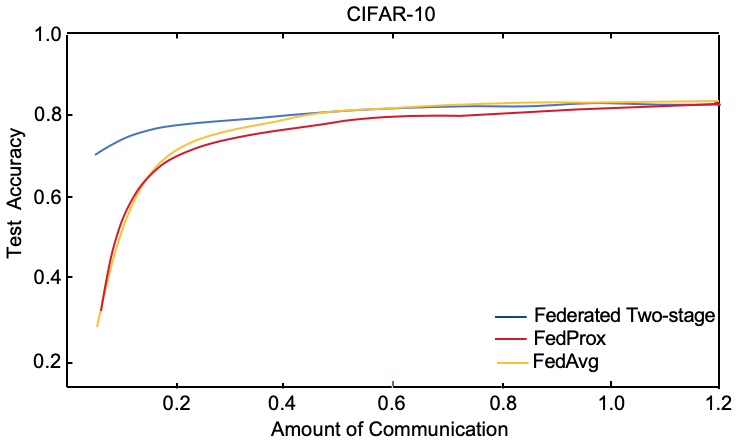}
\includegraphics[width=4.5cm]{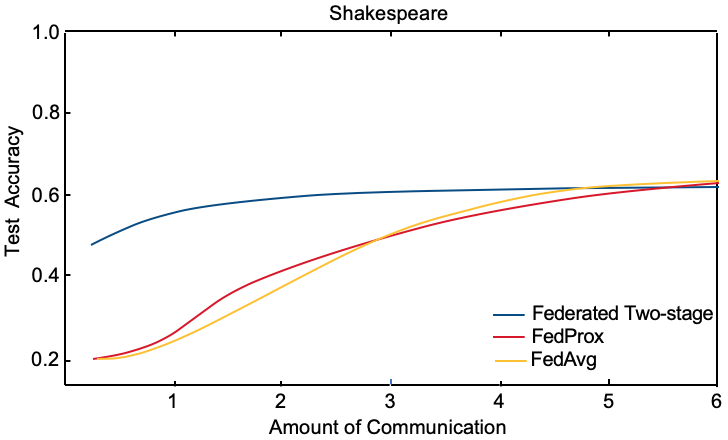}  
\caption{Amount of communicated intermediaries (GB). Federated Two-stage reduce the amount of communicated intermediaries because it only exchanges sign of gradients (1-bit) rather than model parameters (32-bit floating number) and has a smaller global model.}
\label{Figure3}
\end{figure*}

\section{Experiments}
In this section, we first compare the communication efficiency of federated two-stage learning, Federated Averaging and FedProx algorithm \cite{li2018federated}. FedProx algorithm is inspired by FedAvg and aims to tackle heterogeneity in federated networks. It adds a proximal term to the objective that helps improve the stability of the model. The key insight behind FedProx is that it punishes more on the differences between the global and the current model on devices.

Then we design three experiments to evaluate our proposed method. (\romannumeral1) we show that federated two-stage learning with sign-based voting can reduce the communicated intermediaries a lot on MNIST \cite{lecun1998gradient}, CIFAR-10 \cite{krizhevsky2009learning} and Shakespeare \cite{mcmahan2016communication} datasets. By using these datasets, we can compare the efficiency of our method with FedAvg and FedProx algorithm based on their papers' baseline model setting. (\romannumeral2) we evaluate the universality of our method by using data from different distributions (rotated data in CIFAR-10). (\romannumeral3)we show that the designed method is able to block critical information for reconstructing the raw data and prevents data leakage.

In Table \ref{Table2}, $G$ is the gradients size, $m$ is the number of devices, $N_1$ and $n_1$ stand for the number of model parameters and used rounds in proposed method, $N_2$ and $n_2$ refer to the number of model parameters and used rounds in FedAvg and FedProx algorithm. We show the required number of intermediaries that communicated per device as well as the total number of them. In an iteration, sign of gradients is communicated in our system rather than the model parameters in FedAvg and FedProx.

\subsection{Efficiency} 

In the first experiment, we use MNIST, CIFAR-10 and Shakespeare datasets. For MNIST, we partition the dataset in a manner of Non-IID where we first sort the data by label, then divide it into 200 shards of size 300 and give 2 shards to each device. Finally, each device has at most 2 classes, which means the data is in highly non-IID setting. Our local model and hyper-parameters choice just follows the Federated Averaging algorithm paper \cite{mcmahan2016communication}. We use 2 convolution layers as our global model. In addition, we average all trained local models' logits when dealing with new devices prediction in testing process. For CIFAR-10, we partition it into 100 devices and each one contains 500 training and 100 testing examples. This sub-experiment is in balanced and IID setting. We choose a 5-layer network that contains 2 convolution layers with 2 fully connection layers and a layer to produce logits. The global model used still contains 2 convolution layers and testing procedure just follows the former one by averaging all the trained local models' logits when new devices pass their data into the system. Shakespeare dataset is substantially unbalanced because many roles have only a few lines. We filter out the clients with data points less than 10k and get 132 clients in total. Then we partition 80\% of data to training set and amalgamate the left 20\% on a device as test set. The model for this dataset contains 2 LSTM layers which have 256 nodes and we want to predict the next character.

Table \ref{Table1} shows results of efficiency comparison. From 50 runs' averaged results in this experiment, our proposed federated two-stage learning with sign-based voting method maintains similar performance but communicates less($\sim$10$\%$) intermediaries. Our method is more efficient comparing with traditional federated learning algorithms. Key ideas in federated two-stage include (\romannumeral1)We use smashed data which is representative and low-dimension rather than the raw data as input of the global model. (\romannumeral2)Sign-based SGD with the majority vote also helps reduce the communicated intermediaries.

In Figure \ref{Figure2}, we evaluate the performance of Federated Two-stage, FedProx and FedAvg on MNIST, CIFAR-10 and Shakespeare dataset. Our proposed method performs similar results comparing with FedProx and FedAvg but uses less communication rounds. Figure \ref{Figure3} shows the amount of communicated intermediaries using Federated Two-stage, FedProx and FedAvg algorithm. Federated two-stage algorithm exchanges the smallest amount of intermediaries in the system because it only transmits the sign of gradients (1-bit) rather than the model parameters (32-bit floating number) and has a smaller global model comparing with other methods.

\begin{figure}[tb]
\centering
\includegraphics[scale=0.5]{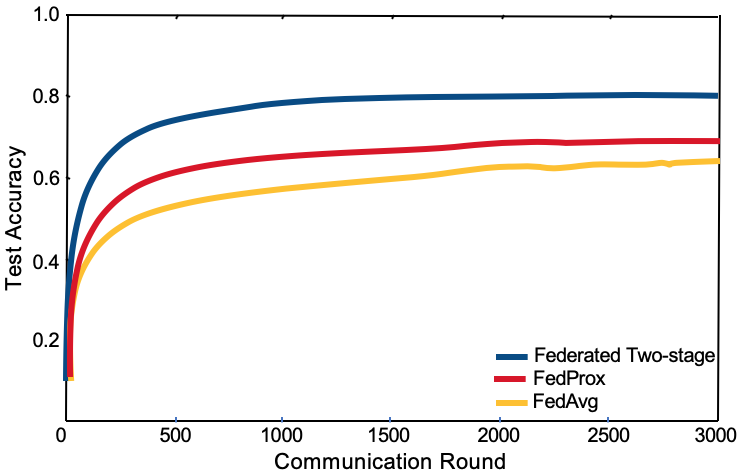}
\caption{Performance comparison between Federated Two-stage, FedAvg and FedProx on the rotated CIFAR-10 dataset. When dealing with a new device with different data distribution, Federated Two-stage performs better than traditional federated learning algorithms}
\label{Figure4}
\end{figure}

\begin{figure*}[tb]
\centering 
\includegraphics[width=2.7cm]{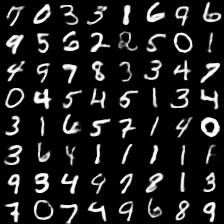}  
\includegraphics[width=2.7cm]{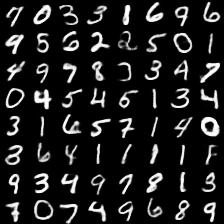}  
\includegraphics[width=2.7cm]{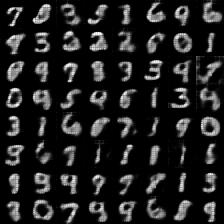}  
\includegraphics[width=2.7cm]{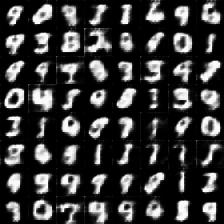}  
\includegraphics[width=2.7cm]{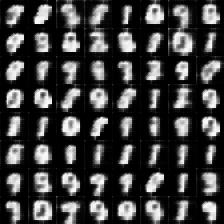}  
\caption{Reconstruction results of smashed data in Federated Two-stage. $\alpha_1$ increases (0.1, 0.3, 0.5, 0.7 and 0.9) from left to right. It is hard to reconstruct from smashed data as $\alpha_1$ increasing because smaller distance correlation blocks more critical information required by reconstruction.}
\label{Figure5}
\end{figure*}

\begin{figure}[tb]
\centering
\includegraphics[scale=0.5]{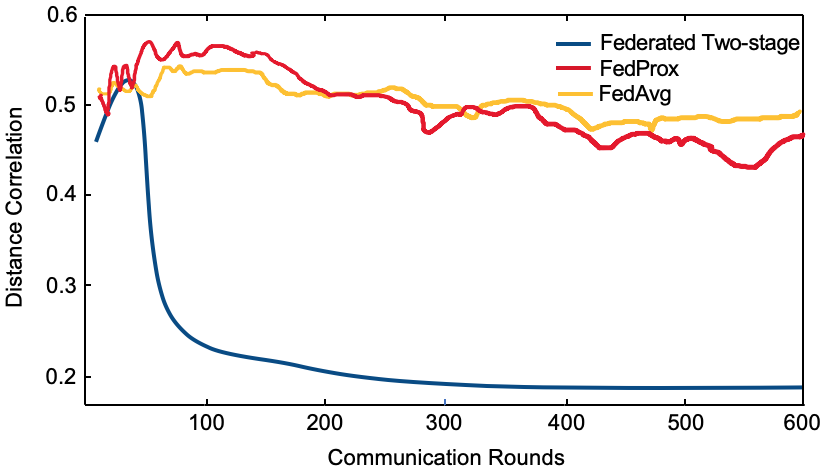}
\caption{Reduction comparison of distance correlation using Federated Two-stage, FedProx and FedAvg algorithm. Our method has smaller distance correlation comparing with other algorithms and protects the system from data leakage.}
\label{Figure6}
\end{figure}

\subsection{Universality} 
In this experiment, we demonstrate that the proposed method can deal with data from new sources or distributions. We spilt the CIFAR-10 dataset into 100 devices. Each device contains 500 training and 100 testing samples. In addition, we introduce a new device with rotated CIFAR-10 examples which contains 1000 training and 200 testing data. The data distribution in this new device is different from the original training dataset. The model architecture contains two convolution layers with two fully connection layers and a transformation layer. Still, the global model is a two convolution layer network. Other hyper-parameters setting follows the Federated Averaging paper.

We first train the model on 100 devices using proposed method, FedAvg and FedProx algorithm. Then the new device is introduced and trains until model converge. Figure \ref{Figure4} shows the accuracy of Federated Two-stage, FedAvg and FedProx algorithm. Our proposed method outperforms the traditional model because federated two-stage learning helps model learn representative patterns in different data distributions by using smashed data.

\subsection{Privacy Preserving} 
We design this experiment to evaluate the potential data leakage risk in our federated two-stage learning with sign-based voting framework. In the third section we discussed potential data leakage in communicated intermediaries and analyzed this risk theoretically. We prove the connection between distance correlation and Kullback-Leibler divergence. Minimizing the logarithm of distance correlation between the smashed data and the raw data, we minimize Kullback-Leibler divergence which is a measure of invertibility of the smashed data. In this experiment, we show that our proposed architecture reduces this data leakage risk a lot on MNIST dataset. The local model used in the experiment contains a CNN with two convolution layers, a fully connection layer and a transformation layer. The global model still contains two convolution layers. All hyper-parameters follow former experiments. Figure \ref{Figure6} shows the reduction in distance correlation between the smashed data and the raw data using Federated Two-stage, FedProx and FedAvg algorithm. Because FedProx and FedAvg do not contain the cut layer, we use representations generated by the first convolution layer of model. Federated two-stage has smaller distance correlation comparing with other algorithms in training and protects the system from data leakage better. Even in the first few rounds, the distance correlation is below 0.55 which means the data leakage risk is relatively small.

In the next sub-experiment, we show reconstruction results from the smashed data in our system. Smashed data is informative and low-dimension representations of the raw data, so decoder can reconstruct the smashed data back to the raw data. From equation (\ref{E6}), we can block the critical information that reconstruction needs using larger $\alpha_1$. 

Figure \ref{Figure5} shows reconstruction results from the smashed data on MNIST dataset in different $\alpha_1$. It demonstrates that larger $\alpha_1$ blocks more reconstruction information and builds a more secure model. However, from equation (\ref{E8}), all of three hyper-parameters $\alpha_1$, $\alpha_2$ and $\lambda$ control model together which means larger $\alpha_1$ may not lead to a better model.

\section{Conclusion}

In this paper, we propose a federated two-stage learning with sign-based voting framework that augments prototypical federated learning with a cut layer on local devices and uses sign-based stochastic gradient descent with the majority vote method on model updates. Our method maintains competitive performance while reducing communicated intermediaries, showing great universality across different data sources and protecting model from data leakage. Both empirical and theoretical analysis show that the novel mechanism offers an efficient and privacy preserving scheme which suits for general applications. 

\section*{Broader Impact}
While federated two-stage learning offers a number of potential benefits, providing stronger guarantees in efficiency and privacy is still an emerging direction for future researchers. We hope our insights can inspire future work on federated learning research, which is one of the most important solutions for data privacy concerns in the society.

\end{document}